\documentclass[a4paper,12pt]{article}
\usepackage[utf8]{inputenc}
\def\dis{\displaystyle}
\def\del{\partial}
\def\noi{\noindent}

\def\d{\rm d}

\title{Operator Ordering Ambiguity and Third Quantization}
\author{\small  Yoshiaki Ohkuwa$^1$, Yasuo Ezawa$^2$, Mir Faizal$^3$, \\ \\ \small %\tiny
$^1$Division of Mathematical Science,  Department of Social Medicine,  \\ \small Faculty
of Medicine, %\\ \tiny 
University of Miyazaki,  \\ \small Kihara 5200,   Kiyotake-cho, Miyazaki, 
889-1692, Japan \\  \small  %\tiny 
$^2$Department of Physics, Ehime university, \\ \small 2-5 Bunkyo-cho, Matsuyama, 
790-8577, Japan \\  \small %\tiny
$^3$Department of Physics and Astronomy, 
University of Lethbridge, \\ \small  Lethbridge, AB T1K 3M4, Canada,   \\ \small %\tiny 
Irving K. Barber School of Arts and Sciences,  \\ \small University of British 
Columbia - Okanagan,\\ \small  %\tiny  
Kelowna, British Columbia V1V 1V7, Canada}
\date{}
\begin{document}

\maketitle

\begin{abstract}
In this paper, we  will  constrain the operator ordering ambiguity of Wheeler-DeWitt equation 
by analyzing the quantum fluctuations  in the universe.
This will be done using a third quantized formalism. 
It is expected that the early stages of the universe are dominated 
by quantum fluctuations. 
Furthermore, it is also expected that these quantum fluctuations get suppressed 
with the expansion of the universe.  
We will show that this desired   behavior  of quantum fluctuations  
could  be realized by   a wide ranges of the factor ordering parameters. 
We will examined two different cosmological models, and observe that    
  a similar range of factor ordering parameters  produces this desired behavior in both 
  those cosmological models. 
\end{abstract}

\vspace{3mm}

\section{Introduction} 

The information about the quantum state of the universe can be obtained from
 the wave function of the universe \cite{Hartle83}-\cite{t2}. 
The wave function of the universe can be viewed as a  solution to the  
  Wheeler-DeWitt equation  \cite{DeWitt67}-\cite{Wheeler57}. 
However, there are serious problems 
  with the interpretation of quantum cosmology 
  \cite{mini12}-\cite{Isham}. 
The Wheeler-DeWitt equation is a hyperbolic second order differential equation, 
so that the square of the absolute value of 
the wave function of the universe cannot  be interpreted as the probability density.
This problem  is analogous to the problem which occurs  in the Klein-Gordon equation. However, the 
problem with Klein-Gordon equation can  be resolved by  second quantizing the Klein-Gordon equation. 
 There are several other problems with first quantization, which are resolved by using second quantization. 
 So, just as several problems with the first quantization are resolved by going to second 
  quantization, it has been proposed that third quantization will resolve several problems associated with the second 
  quantized Wheeler-DeWitt equation  \cite{Isham}-\cite{th1}.  
Third quantization is basically a quantum field theory of 
  geometries in the superspace. Thus, in third quantized gravity, the creation and annihilation operators create and annihilate 
  geometries. So, it is possible to study the creation of universe using third quantization \cite{universe}-\cite{universe1}. 
  As the third quantization of gravity can create and annihilate geometries, it is possible to use third quantization to study 
  multiverse \cite{multi}-\cite{multi1}. It has been demonstrated that in such a theory,  
  the quantum state of the multiverse is consistent with  standard cosmological boundary conditions.
  The quantum state of such a multiverse is  found to be squeezed, and can be related to accelerating universes.
  Recently, it has been argued that the third quantization can be used to study the evolution of the
physical constants in   classically disconnected universes, which are 
  quantum-mechanically entangled \cite{dyna}.  Thus, third quantized gravity is an important approach to quantum gravity. 

  It may be noted that third quantization has been been applied in various approaches to quantum gravity. 
  The studies in loop quantum gravity, have led to the development of group 
  field theory \cite{gft12}-\cite{gft14}, 
  and group field cosmology  \cite{gfc12}-\cite{gfc14}, both of which are third quantized theories. 
  Even the third quantization of string theory has been used  to  to properly 
  analyze different aspects of the string theory, and this third quantized string theory is called 
  as the string field theory \cite{Siegel:1988yz}-\cite{fieldt}. The third quantization has been used to 
  analyses the  transitions of a string vacuum state to a cosmological solution  \cite{st}. This was done by 
  analyzing the creation of a pair of two universes from a string vacuum state. As third quantization 
  has been used in various different approaches to quantum gravity,
  the study of third quantization is a very important in  quantum gravity. 
  
It may be noted that third quantization of modified theories of gravity has also been analyzed. 
 The third quantization of 
   Brans-Dicke theories    \cite{ai},  $f(R)$ gravity theories \cite{f}-\cite{f1}
   and   Kaluza-Klein theories     
  \cite{ia} has been studied. It is important to study the suppression of quantum fluctuation in such cosmological models. 
  The quantum uncertainty in third quantization has been studied, and it has been observed that such the quantum fluctuations 
  are suppressed during expansion of the universe \cite{un}-\cite{un1}. 
  Thus, at the beginning of the universe, quantum fluctuations dominate, but they are suppressed as the universe expands. 
  It has been demonstrated that this behavior occurs only for certain values of the factor ordering  parameter \cite{OhkuwaFaizalEzawa1}.
%  In this paper, we will generalize these results to obtain a range of values 
%  for factor ordering parameters. Furthermore, we will analyze two different 
%cosmological models, and observe that they have similar ranges 
%  for the factor ordering parameter. This can indicate the model independence of such results. 
In this paper, we will generalize these results to obtain a range of values for the factor 
ordering parameter, which satisfy this desired behavior. 
We will analyze two cosmological models, and observe that they have similar ranges for the factor ordering parameter. 

In section 2, we review the formulation of the third quantized gravity and apply it to 
the  universe which is filled by a cosmological constant.  
In section 3, quantum fluctuations of the universe will be investigated using the uncertainty principle.  
In section 4, the ranges of the factor ordering parameter will be calculated, which satisfy  
the desired behavior. 
In section 5, another cosmological model  will be studied, to investigate the possibility of  
model dependence of the above range of factor order parameters.   
In section 6, we will summarize our results.

 \section{Third Quantized Theory}

    In order to analyze the third quantization of cosmological models, we need 
 to identify the scale factor of the universe with a 'time' parameter for this third quantized system. 
Then we would expect that the quantum fluctuations would be suppressed at late times,
 and the universe would be described by a  classical geometry. 
However,  at the beginning of the universe quantum fluctuations would dominate. 
This requirement can be used to constrain the operator ordering ambiguity of 
the Wheeler-DeWitt equation \cite{OhkuwaFaizalEzawa1}. 
In fact, such quantum fluctuations for a geometry can be analyzed in the third quantized
 formalism using the uncertainty principle \cite{un}-\cite{un1}. 

Now let the Wheeler-DeWitt equation   be given by 
$
 H \psi(h, \phi) =0,  
$
where $h$ is the induced three metric, $\phi$ is the value of the matter field on the
 boundary, and $H$ is the Hamiltonian constraint obtained from general relativity
 \cite{DeWitt67}-\cite{Wheeler57}.
Then we can write the third quantized Lagrangian for this system as 
$\bar{\mathcal{L}}_{3Q} =   \varphi (h, \phi) H \psi(h, \phi) 
$. 
When this system is quantized we will obtain creation $b^\dagger$ and
 annihilation operators $b$, such that for vacuum state $|0>$, we would have 
$ b |0> =0. $ These creation and annihilation operators will create and annihilate geometries.  
We have used $b$, for the annihilation operator to distinguish  it 
from the scalar factor of the universe, which is denote by $a$. 
Now for specific minisuperspace models, we can  identify the scale factor of the universe
 $a$, 
with the time of this quantum system \cite{mini12}-\cite{mini14}.   
So, when this scale factor is small quantum fluctuations should dominate this system, and
 when this scale factor is large the quantum fluctuations should be suppressed.

Now as an example, in the cosmological model, where the universe is filled by a 
cosmological constant \cite{c1}-\cite{c2}, a  flat Friedmann-Lemaitre-Robertson-Walker
 metric can be written as 
$$
ds^2=-dt^2+ a^2 (t) \sum_{k=1}^3 (dx^k)^2.                                    \eqno(2.1)
$$
Here $a(t)$   is the scale factor of the universe,  and $a(t)$ denotes the cosmological
 evolution of this system and also the size of the universe. 
It may be noted that the
 Wheeler-DeWitt equation for this system can be written as (here we set $8 \pi G=1$)
$$
\left[ {1 \over a^{p_o}} {{\d} \over {\d} a}
a^{p_o} {{\d} \over {\d} a} 
+ 12 \Lambda a^4 \right] \psi (a)  = 0.                                     \eqno(2.2)
$$                                            

We observe that there is a factor ordering ambiguity due to the parameter 
$p_0$ in such minisuperspace models
\cite{fo12}-\cite{fo14}.  However,  it has been demonstrated that such factor ordering
 can be constrained by the physics of this system. 
This is because    the quantum fluctuations dominate at the early times and are
 suppressed at the later times, only for certain values of operator ordering parameter  
\cite{OhkuwaFaizalEzawa1}. 
However, it is important to know the  exact ranges of the factor ordering parameter for
 which the universe evolves as desired. 
Furthermore, it is important to know if this result hold for different cosmological models,
 or if it is a model dependent result. 
So, in this paper, we will analyze two different cosmological models, and observe that,
 since these two models have very wide common ranges of $p_o$ which produce the
 correct desirable behavior, there is the possibility that there exists some 
 desirable model independent operator ordering parameter $p_o$ .

Now we can use the formalism of third quantization and write
 the third quantized Lagrangian for this quantum system   \cite{th}-\cite{th1}, 
$$
{\cal L}_{3Q} = {1 \over 2}
\left[  a^{p_o} \left( {{\d}\psi(a) \over {\d} a}\right)^2
-12\Lambda a^{p_o +4} \psi (a)^2
\right] \ .                                                                            \eqno(2.3)
$$ 
Using the standard formalism of third quantization, we can write the third quantized 
 $\rm Schr\ddot{o}dinger$ equation for this system as  \cite{OhkuwaFaizalEzawa1}
$$
\left\{
\begin{array}{ll}
&\dis{i{\del \Psi (a, \psi) \over \del a}} = {\hat {\cal H}}_{3Q} 
\Psi (a, \psi) \ , \\[5mm]
&\qquad\ \ {\hat {\cal H}}_{3Q}= 
\dis{{1 \over 2}\left[- {1 \over a^{p_o}} 
{\del^2 \over \del \psi^2}  
+12\Lambda a^{p_o +4} \psi^2
\right]} \ . 
\end{array}                                                  
\right.                                                                                   \eqno(2.4)
$$
Here we ignored the operator ordering problem in the first term of 
$ {\hat {\cal H}}_{3Q}$  for simplicity. 
Now the  $\Psi (a, \psi )$ is the third quantized wave function of the universes.  
The wave function of the universes  $\Psi (a, \psi )$ can be obtained as a solution to the 
 third quantized $\rm Schr\ddot{o}dinger$ equation, instead of the Wheeler-DeWitt
 equation.

\section{Quantum Fluctuations}\label{2} 
As we have assumed that   the quantum fluctuations are suppressed at later times, 
and dominate at earlier times, it is important to analyze  these quantum fluctuations. 
These quantum fluctuations can be analyzed using the uncertainty principle for these minisuperspace 
models. In this section, we will analyze such quantum fluctuations for a universe filled with the cosmological 
constant.  
The scale factor for such a universe can be identified with the time 
variable which describes the evolution of the quantum 
system and the size of this geometry. So,  we can denote the initial state of this quantum system  by the limit 
$a \rightarrow 0$, and this quantum system is expected to evolve to $a \rightarrow \infty$. 
Now it is expected that quantum fluctuations should dominate the limit $a \rightarrow 0$. 
Furthermore, as the universe at later times is 
represented by a classical geometry, we expect that these quantum fluctuations are suppressed in the 
limit  $a \rightarrow \infty$.

To analyze the  uncertainty for this third quantized quantum system, we first 
assume  a Gaussian form of the  solution  
$$
\Psi (a, \psi) = C {\rm exp} \left\{ -{1 \over 2}A(a)
[\psi-\eta (a)]^2 +i B(a)[\psi-\eta (a)]
\right\} \ ,                                                                           \eqno(3.1)
$$
where  $C$ is a real constant, $A(a) \equiv D(a)+iI(a)$, and  $A(a), B(a), \eta (a)$ 
should be determined from Eq. (2.4). 
It is possible to define an inner product for two third quantized wave functions,  
$\Psi_1$ and $\Psi_2$  as follows, 
$$
\langle \Psi_1 , \Psi_2 \rangle 
=\int^{\infty}_{-\infty} \! d \psi \, \Psi_1^*(a,\psi)
 \Psi_2(a,\psi)          .                                                              \eqno(3.2)
$$
Now we can use this equation to obtain the     uncertainty for this third quantized quantum system. 
This can be done by first writing the  dispersion of $\psi$ as 
$$
(\Delta \psi)^2 \equiv \langle \psi^2 \rangle
-\langle \psi \rangle^2 \ , \qquad
\langle \psi^2  \rangle 
= {\langle \Psi ,  \psi^2 \Psi \rangle \over 
\langle \Psi , \Psi \rangle } \ .                                                    \eqno(3.3)
$$
Furthermore, we can also write  the dispersion of $\pi$ as 
$$
(\Delta \pi)^2 \equiv \langle \pi^2 \rangle
-\langle \pi \rangle^2 \ , \qquad
\langle \pi^2  \rangle 
= {\langle \Psi ,  \pi^2 \Psi \rangle \over 
\langle \Psi , \Psi \rangle } \ .                                                  \eqno(3.4)
$$
We can write the  
the  uncertainty for these geometries as  \cite{OhkuwaFaizalEzawa1}
$$
(\Delta \psi)^2 (\Delta \pi)^2
={1 \over 4} \Biggl( 1+ {I^2 (a) \over D^2 (a)} 
\Biggr) \  .                                                                            \eqno(3.5)
$$

It may be noted that the equation for  $A(a)$ can be written  
as 
$$
-{i \over 2}{{\d} A(a) \over {\d} a}
=-{1 \over 2 a^{p_o}} A(a)^2 
+ 6 \Lambda a^{p_o +4} \ .                                                     \eqno(3.6)
$$
This equation for $A(a)$  is sufficient to obtain the  uncertainty in geometry as  $A(a)=D(a)+iI(a)$.

This cosmological model with      $p_o \neq 1$ has been studied 
\cite{OhkuwaFaizalEzawa1}. 
Now if we define 
$$
z \equiv {2 \sqrt{{ \Lambda} \over 3}} a^3 \ ,                              \eqno(3.7)
$$ 
we obtain  the following solution for $A(z)$,  
$$
A(z) =  \dis{-i \, 6\sqrt{\Lambda \over 3}
\left( {z \over 2\sqrt{\Lambda \over 3}} \right)^{p_o+2 \over 3}
{c_J J_{-5-p_o \over 6} (z) + c_Y Y_{-5-p_o \over 6} (z) \over 
c_J J_{1-p_o \over 6} (z) + c_Y Y_{1-p_o \over 6} (z)} }
\ ,                                                                                       \eqno(3.8)
$$
where $J_{\nu}$ and $Y_{\nu}$ are Bessel functions of order $\nu$ and 
$c_J$ and $c_Y$ are arbitrary complex constants. Now using this equation, it is possible to 
 obtain  both $D$ and $I$. 

 Now assuming $c_J c^*_Y - c^*_J c_Y \neq 0$, 
we get \cite{OhkuwaFaizalEzawa1}
$$
\begin{array}{ll}
\dis{I(z)^2 \over D(z)^2}
=&\!\!\!\dis -{\pi^2 z^2 \over 4 (c_J c^*_Y - c^*_J c_Y)^2} \\[6mm]
&\!\!\!\!\!\!\times
\biggl[ 2\vert c_J \vert^2 J_{-5-p_o \over 6}(z) J_{1-p_o \over 6}(z)
+2\vert c_Y \vert^2 Y_{-5-p_o \over 6}(z)Y_{1-p_o \over 6}(z)  \\[3mm]
&%\quad 
\!\!\!+(c_J c_Y^* + c_J^* c_Y)
\Bigl( J_{-5-p_o \over 6}(z) Y_{1-p_o \over 6}(z) %\\[3mm]
%&\qquad \qquad \qquad \quad \quad
+ J_{1-p_o \over 6}(z) Y_{-5-p_o \over 6}(z) \Bigr)
 \biggr]^2 \ . 
\end{array}                                                                           \eqno(3.9)
$$
Thus, the uncertainty of the quantum system can be obtained. So, we can now 
 use the requirements for quantum fluctuations to constrain the ranges 
of the factor ordering operator for this quantum system.

\section{Operator Ordering}\label{b}
Now we can analyze specific ranges of the operator ordering parameter for 
this cosmological model. It may be noted that as  this quantum system evolves to 
$a \rightarrow \infty$, it also evolves to 
$z \rightarrow \infty$, and in this limit, we have 
\cite{Abramowitz-Stegun}
$$
J_{\nu} (z) \sim \sqrt{ 2 \over \pi z} 
\cos \left( z-{\nu \pi \over 2} - {\pi \over 4} \right) \ , 
\quad 
Y_{\nu} (z) \sim \sqrt{ 2 \over \pi z} 
\sin \left( z-{\nu \pi \over 2} - {\pi \over 4} \right) 
\ ,                                                                                         \eqno(4.1)
$$
where $\nu = {-5-p_o \over 6} \ {\rm and} \ {1-p_o \over 6}$. 
Now we can also  write  
$$
\begin{array}{ll}
\dis{I(z)^2 \over D(z)^2}
&\sim -\dis{1 \over  (c_J c^*_Y - c^*_J c_Y)^2} \\[6mm]
&\qquad\times
\biggl[ 2\vert c_J \vert^2 
\cos \Bigl( z + {p_o + 2 \over 12}\pi \Bigr)
\cos \Bigl( z + {p_o -4 \over 12}\pi \Bigr) \\[5mm]
&\qquad \ \, 
+2\vert c_Y \vert^2 \sin \Bigl( z + {p_o + 2 \over 12}\pi \Bigr)
\sin \Bigl( z + {p_o -4 \over 12}\pi \Bigr)  \\[5mm]
&\qquad \ \, 
+(c_J c_Y^* + c_J^* c_Y)
\sin \Bigl( 2z + {p_o -1 \over 6} \pi \Bigr)
 \biggr]^2 \\[5mm]
&\sim O(1)    \ . 
\end{array}                                                                               \eqno(4.2)
$$ 
Thus, as $a \rightarrow \infty$,  we obtain a classical geometry, 
and this occurs as the quantum fluctuations are suppressed in this limit. 

Now the initial state of this quantum system will be denoted by 
$a \rightarrow 0$, and this also corresponds to $z \rightarrow 0$. 
It is important to analyze the   ranges of $p_o$ for which    the   uncertainty 
becomes of order one, and the ranges for which it tends to infinity. 
The uncertainty of order one corresponds to a classical geometry, and the uncertainty 
of order infinity corresponds to a state for which the geometry is dominated by quantum fluctuations.  
 
Now we  simplify the notation and define,  
$$
\nu_1={1-p_o \over 6 } , \quad  \nu_2={-5-p_o \over 6}\  ; \quad  \nu_1 = \nu_2 + 1 . 
                                                                                           \eqno(4.3)
$$
So, we  consider  the limit $z \rightarrow 0$ , and    use  the 
relations \cite{Abramowitz-Stegun}
$$
\left\{\begin{array}{ll}
\dis{ J_{\nu}(z)} 
&\sim \dis{ {1 \over \Gamma ( {\nu +1} )}
 \left( {z \over 2} \right)^{\nu} \quad (\nu \neq -1, -2, -3, \cdots )
} , \\[6mm]
J_{-n} (z) &= (-1)^n J_n (z) , \quad Y_{-n} (z) =  (-1)^n Y_n (z) 
\quad  ( n = 1,2,3, \cdots )  
 , \\[6mm]
\dis{Y_{0} (z)} 
&\sim \dis{ {2 \over \pi} \ln z , 
\qquad Y_{\nu}  (z)
\sim - {1 \over \pi} \Gamma (\nu)  \left( {z \over 2} \right)^{- \nu}
\quad ( {\rm Re}\ \nu > 0 )  
},  
\end{array}\right.                                                                    \eqno(4.4)                         
$$
along with  
$$
\dis{ Y_{\nu} (z) = {J_{\nu} (z) \cos ( \nu \pi ) - J_{-\nu} (z) 
\over  \sin (\nu \pi ) } \quad ( \nu \neq {\rm integer} ).
}                                                                                         \eqno(4.5)
$$
Now,  we  divide the ranges of $\nu_1 , \ \nu_2$  as 
\\[2mm]  
\noi 1) $\nu_1 = 0$ or $\nu_2 = 0$, \ 
2) $\nu_1 >0 , \ \nu_2 >0$, \ 
3) $\nu_1 >0 , \ \nu_2 <0$, \ 
4) $\nu_1 <0 , \ \nu_2 <0$. 
\\[2mm]
It may be noted  that as  $\nu_1 = \nu_2 +1$, we do not need to consider 
$\nu_1 <0 , \ \nu_2 >0$.

Let us first   consider the case, when $\nu_1 = 0$ or $\nu_2 = 0$. We first note that 
$\nu_1=0$,  implies  $p_o = 1$, and as we have assumed $p_o \neq 1$, 
we can omit this case. So, now    
$\nu_2=0$, implies  $p_o = -5$ and $\nu_1 = 1$. 
Now when  $z \rightarrow 0$, we can write 
$$
\left\{\begin{array}{ll}
\dis{ J_{0}(z)} 
&\sim \dis{ 1 ,
\qquad J_{1}(z) 
\sim { \left({z \over 2} \right) \over \Gamma (2) } \rightarrow 0
 },  \\[6mm]
\dis{Y_{0}(z)} 
&\sim \dis{ {2 \over \pi} \ln z \rightarrow - \infty , 
\qquad Y_{1}(z) 
\sim - {\Gamma (1) \over \pi} 
\left({z \over 2} \right)^{-1} \rightarrow - \infty
} .  
\end{array}\right.                                                              \eqno(4.6)                         
$$
The  largest term in Eq. (3.9) for this case is proportional to $Z$, where 
$$
Z = z^2 [ Y_0 (z) Y_1 (z) ]^2 \sim {16 \over \pi^4} 
( \ln z)^2 \rightarrow \infty.                                                  \eqno(4.7)
$$
So,  when $p_o = -5$, we obtain  
$$
\Delta \psi \cdot \Delta \pi \rightarrow \infty \quad  
( z \rightarrow 0 ) .                                                            \eqno(4.8)
$$

Now let us consider the case, when $\nu_1 >0,$ and  $\nu_2 >0$.
In this case, we can again  write  the largest term in Eq. (3.9) 
 proportional to $Z$, where 
$$
Z = z^2 [Y_{\nu_1} (z) Y_{\nu_2} (z) ]^2 \sim 
\left[ {1 \over \pi^2} \Gamma (\nu_1) \Gamma (\nu_2) 
\left( {1 \over 2} \right)^{-(\nu_1 + \nu_2)} \right]^2 
z^{2-2(\nu_1 + \nu_2)}. 
                                                                                        \eqno(4.9)
$$
Now for $\nu_1 >0, $ and $ \ \nu_2 >0$ implies $p_o < -5$, and so 
we obtain  $2-2(\nu_1 + \nu_2) < 0$ . 
This term becomes infinity when $z \rightarrow 0$. 
For  $p_o < -5$, we also obtain  
$$
\Delta \psi \cdot \Delta \pi \rightarrow \infty \quad  
( z \rightarrow 0 ).                                                              \eqno(4.10)
$$

Let us also consider  $\nu_1 >0, $ and $\ \nu_2 <0$. 
This case implies $0 < \nu_1 < 1 , \ -1 < \nu_2 < 0 $, and we know that 
$\nu_1$ and  $\nu_2$ are not integer.  
Now when  $z \rightarrow 0$,  we obtain 
$$
\left\{\begin{array}{ll}
\dis{ J_{\nu_1}(z)} %[&]
\sim \dis{ {1 \over \Gamma ( \nu_1 +1)}
 \left( {z \over 2} \right)^{\nu_1} \rightarrow 0 ,
\qquad J_{\nu_2}(z) 
\sim {1 \over \Gamma ( \nu_2 +1)}
\left({z \over 2} \right)^{\nu_2} \rightarrow \infty
 },  \\[6mm]
\dis{Y_{\nu_1}(z)} %[&]
\sim \dis{ -{1 \over \pi} \Gamma (\nu_1) 
\left( {z \over 2} \right)^{-\nu_1} \rightarrow - \infty , 
\qquad J_{-\nu_2}(z) 
\sim {1 \over  \Gamma ( - \nu_2 +1)} 
\left({z \over 2} \right)^{-\nu_2} } \\[6mm]
\qquad\qquad\qquad\qquad\qquad\qquad\qquad\qquad\qquad\qquad \ \ 
\rightarrow 0 ,  
\end{array}\right.                                                                 \eqno(4.11)                         
$$
and we also obtain 
$$
Y_{\nu_2} \sim \left\{
\begin{array}{ll}
&\dis{ {\cos (\nu_2 \pi) \over \sin (\nu_2 \pi)}
{1 \over \Gamma (\nu_2 +1)}  \left( {z \over 2} \right)^{\nu_2}
\rightarrow + \infty \quad \left( -1< \nu_2 <-{1 \over 2} \right) 
 },  \\[6mm] 
&\dis{ {\cos (\nu_2 \pi) \over \sin (\nu_2 \pi)}
{1 \over \Gamma (\nu_2 +1)}  \left( {z \over 2} \right)^{\nu_2}
\rightarrow - \infty \quad \left( -{1 \over 2}< \nu_2 <0 \right) 
 },  \\[6mm] 
&\dis{ -{J_{1 \over 2} (z) \over \sin \left( - {\pi \over 2} \right) }
\sim  {1 \over \Gamma \left( {3 \over 2} \right)} 
 \left( {z \over 2} \right)^{1 \over 2}
\rightarrow 0 \qquad 
\left( \nu_2 = - {1\over 2} \right)
} .  
\end{array}   \right.                                                              \eqno(4.12)                         
$$
Thus, the term which is the largest in Eq. (3.9) in this case is 
  proportional to $Z_1, Z_2, Z_3$, such that  
$$
\begin{array}{ll}
Z_1 = z^2 [J_{\nu_2} (z) Y_{\nu_1} (z) ]^2 
&\dis{ \sim \left[ {1 \over \Gamma (\nu_2 +1)}
\left( - {\Gamma (\nu_1) \over \pi} \right) 
\left( {1 \over 2} \right)^{-\nu_1 +\nu_2} \right]^2 
z^{2+2(- \nu_1 + \nu_2)} 
}  \\[6mm]
&\sim O(1)  . 
\end{array}                                                                            \eqno(4.13)
$$
$$
\begin{array}{ll}
Z_2 = z^2 [Y_{\nu_1} (z) Y_{\nu_2} (z) ]^2 
&\dis{ \sim \left[  - {\Gamma (\nu_1) \over \pi}  
{\cos (\nu_2 \pi) \over \sin (\nu_2 \pi)}
{1 \over \Gamma (\nu_2 +1)}
\left( {1 \over 2} \right)^{-\nu_1 +\nu_2} \right]^2 
} \\[6mm]
&\quad \times z^{2+2(- \nu_1 + \nu_2)} 
  \\[6mm]
&\sim O(1) .
\end{array}                                                                            \eqno(4.14)
$$
$$
\begin{array}{ll}
Z_3 = &z^2J_{\nu_2} (z) Y_{\nu_1} (z) Y_{\nu_1} (z) Y_{\nu_2} (z)  \\[6mm] 
&\dis{ \sim \left( {1 \over \Gamma (\nu_2 +1)} \right)^2
\left( - {\Gamma (\nu_1) \over \pi} \right)^2 
{\cos (\nu_2 \pi) \over \sin (\nu_2 \pi)}
\left( {1 \over 2} \right)^{-2\nu_1 +2\nu_2} 
z^{2+2(- \nu_1 + \nu_2)} 
}  \\[6mm]
&\sim O(1) .
\end{array}                                                                           \eqno(4.15)
$$
Here we have used, 
$2 + 2(-\nu_1 +\nu_2) = 0 $. It may be noted that   
  the case $\nu_2 = - 1/2$ has not been considered in  Eqs. (4.14) and  (4.15). 
Now $0< \nu_1 <1$ implies  $-5 < p_o < 1$, and so 
  for  $-5 < p_o < 1$, we obtain   
$$
\Delta \psi \cdot \Delta \pi \rightarrow O(1) \quad  
( z \rightarrow 0 ).                                                                \eqno(4.16)
$$

Now let us consider the values  $\nu_1 <0 , $ and $  \nu_2 <0$. 
To analyze the initial state of the quantum system for this case, we need to analyze the 
  behavior of 
Bessel functions in the limit $ z \rightarrow 0$. 
Now  we can  write $\nu = \nu_1\  {\rm or}\  \nu_2$, 
so we can also write $\nu < 0$. 
When  $ z \rightarrow 0$, we can write 
$$
\left\{\begin{array}{lcl}
J_{\nu} (z) &\sim& \dis{ {1 \over \Gamma (\nu +1)} 
\left( {z \over 2} \right)^{\nu}
\rightarrow \infty 
\quad ( \nu \neq -1, -2, -3, \cdots ) 
} , \\[6mm]
J_{-\nu} (z) &\sim& \dis{ {1 \over \Gamma (-\nu +1)} 
\left( {z \over 2} \right)^{-\nu}
\rightarrow 0 
} ,  
\end{array}\right.                                                                  \eqno(4.17)
$$
We can also write 
$$
J_{-n} (z) = (-1)^n J_n (z) \sim \dis{ 
(-1)^n {1 \over \Gamma (n+1) } \left( {z \over 2} \right)^n 
\rightarrow 0 \quad ( n= 1, 2, 3, \cdots ) 
}.                                                                                     \eqno(4.18)
$$
Using the relation \cite{Moriguchi}
$$
\Gamma (\nu) \Gamma (1-\nu) = { \pi \over \sin (\pi \nu) },                           \eqno(4.19)
$$    we observe that as  $ z \rightarrow 0$,  
$$
\dis{ Y_{\nu} (z) \sim 
{\cos (\nu \pi) \over \sin (\nu \pi)}
{1 \over \Gamma ( \nu +1 )}
\left( {z \over 2} \right)^{\nu} 
= \cos (\nu \pi) {\Gamma (1- \nu) \over \nu \pi}
\left( {z \over 2} \right)^{\nu} 
}
$$
$$
\rightarrow \left\{ 
\begin{array}{ll}
- \infty &\quad ( -{1 \over 2 } < \nu <0 )  \\[6mm]
+ \infty &\quad (-2n+{1 \over 2} < \nu < -2n+1, \ 
-2n+1 < \nu < -2n+{3 \over 2} )  \\[6mm]
- \infty &\quad (-2n-{1 \over 2} < \nu < -2n, \ 
-2n < \nu < -2n+{1 \over 2} ) 
\end{array} \right.                                                               \eqno(4.20)
$$
$$
\dis{ Y_{-n} (z) = (-1)^n Y_n (z)
\sim (-1)^{n+1} {\Gamma (n) \over \pi}
\left( {z \over 2} \right)^{-n}
\rightarrow (-1)^{n+1} \infty
}                                                                                       \eqno(4.21)
$$ 
$$
\dis{ Y_{-n+{1 \over 2}} (z) 
= - { J_{n-{1 \over 2}} (z) \over \sin \left( \left( -n+{1 \over 2}\right) \pi \right) }
= (-1)^{n+1} J_{n-{1 \over 2}} (z)  \rightarrow 0
},                                                                                     \eqno(4.22)
$$
where $n=1, 2, 3, \cdots$ . 

Using the above relations and Eq. (4.3), it is seen that the terms in Eq. (3.9) 
which could be large for this case,  
include the terms that are proportional to  $Z_1, Z_2$. For $Z_1$, we note that  
$$
\begin{array}{ll}
Z_1 = z^2 [ J_{\nu_1} (z) J_{\nu_2} (z) ]^2
&\sim \dis{ 
\left[ { \left( {1 \over 2} \right)^{\nu_1 + \nu_2}
\over \Gamma (\nu_1 +1) \Gamma (\nu_2 +1) } \right]^2 
z^{2+2( \nu_1 + \nu_2 )}
\rightarrow \infty} \\[6mm]
&\qquad\qquad\qquad\qquad\qquad\qquad
(1< -\nu_1-\nu_2) .   
\end{array}                                                                              \eqno(4.23)
$$
Here we have omitted the  case when  $\nu_1$ or $\nu_2$ is a   
negative integer, namely 
$p_o = 6n+1 \ (n=1, 2, 3, \cdots )$. 
Since $\nu_1 <0, \ \nu_2 <0$, this implies that  $p_o > 1$, and  $1< -\nu_1-\nu_2$ holds. 
Thus,  this $Z_1$ can  becomes infinity, 
when $p_o > 1$ and $p_o \neq 6n+1 \ (n=1, 2, 3, \cdots )$. For $Z_2$, we note that    
$$
\begin{array}{ll}
Z_2 = z^2 [Y_{-n_1} (z) Y_{-n_2} (z) ]^2 
&\sim \dis{ \left[ 
{\Gamma (n_1) \Gamma (n_2) \over \pi^2} 
\left( {1 \over 2} \right)^{-n_1 -n_2} \right]^2 
z^{2-2(n_1 +n_2)} } \\[6mm]
&\rightarrow \infty \quad
(n_1 =1, 2, 3, \cdots ; \ n_2 = n_1 +1 )  \ , 
\end{array}                                                                         \eqno(4.24)
$$
where $ \nu_1 = -n_1 , \  \nu_2 = -n_2 $. 
As $n_1 =1, 2, 3, \cdots$ implies that 
$p_o = 6n_1 +1$, this term   becomes infinity, 
when $p_o=6n +1 \ (n=1, 2, 3, \cdots )$. 
So, we obtain that  Eq. (3.9)  becomes infinity for both these cases, 
when $p_o > 1$. 
Therefore, we observe that  when  $p_o > 1$,  
$$
\Delta \psi \cdot \Delta \pi \rightarrow \infty \quad  
( z \rightarrow 0 ).                                                               \eqno(4.25)
$$

Let us summarize above consideration. 
We obtain that 
$$
{\rm when} \ \  p_o > 1 \  {\rm or} \ p_o \leq -5 \ , \quad 
\Delta \psi \cdot \Delta \pi \rightarrow \infty \quad  
( z \rightarrow 0 ) ,                                                                   \eqno(4.26)
$$
which means that when $p_o > 1$ or $p_o \leq -5$ the quantum fluctuations  dominate 
the universe at the early times.  
On the other hand we obtain that   
$$
{\rm when} \ -5 < p_o < 1 \ , \quad 
\Delta \psi \cdot \Delta \pi \rightarrow O(1) \quad  
( z \rightarrow 0 ) ,                                                                   \eqno(4.27)
$$
which means that when $-5 < p_o < 1$ universe can become classical at the early times. 
Since we expect that the quantum fluctuations dominate the universe at the early times, 
$p_o > 1$ or $p_o \leq -5$ is desirable. 
Note that we have assumed $p_o \neq 1$ , and above result is consistent with 
Ref.  \cite{OhkuwaFaizalEzawa1}.

\section{Model Dependence}\label{a}
It is important to analyze if these desirable values of factor ordering depend on a specific cosmological 
model, or if they are model independent.  So, in this section, we will perform a similar analysis for a different 
cosmological model. In this cosmological model,  a closed universe  is filled
 with a constant vacuum energy density  $\rho_v$ and radiation $\epsilon$, and the 
  Wheeler-DeWitt equation for this model can be written as  \cite{d}-\cite{fd}
$$
\left[ {{\d}^2 \over {\d} a^2}
+ {p_o \over a} {{\d} \over {\d} a} 
- k_2 a^2 + k_4 \rho_v a^4 + k_0 \epsilon
 \right] \psi (a)  = 0,                                    \eqno(5.1)
$$                                                
where $a$ is  the scale factor for this closed  universe,    
$p_o$ is the operator ordering parameter for this cosmological model, and 
$$ 
k_2={9 \pi^2 \over 4G^2 \hbar^2} \ , \ 
k_4={6 \pi^3 \over G \hbar^2} \ ,  \ 
k_0={6 \pi^3 \over G \hbar^2} 
\ .  \                                                      \eqno(5.2)
$$   
It may be noted that the  wave function of the universe for this cosmological model has
 been discussed,  so we can perform the above analysis for this cosmological model 
 \cite{fd}.

Now in this cosmological model, we again assume a Gaussian form   of solution 
for the third quantized $\rm Schr\ddot{o}dinger$ equation. So, 
 uncertainty in its geometry can  also be obtained using the same formalism 
 \cite{OhkuwaFaizalEzawa1}. 
In this reference we found that at the late times for any $p_o$ the universe becomes classical, 
since the quantum fluctuation becomes minimum. 
Now at the early times for  $p_o \neq 1$, we can write  
$$
z \equiv \sqrt{k_0 \epsilon} \  a.                                                        \eqno(5.3)
$$
So, initial state for this quantum system can be written as   $a \rightarrow 0$, and this
 also corresponds to $z \rightarrow 0$. 
For this initial state, we obtain \cite{OhkuwaFaizalEzawa1}, 
$$
\begin{array}{ll}
\dis{I(z)^2 \over D(z)^2}
=&-\dis{\pi^2 z^2 \over 4 (c_J c^*_Y - c^*_J c_Y)^2} \\[6mm]
&\times
\biggl[ 2\vert c_J \vert^2 J_{-1-p_o \over 2}(z) J_{1-p_o \over 2}(z)
+2\vert c_Y \vert^2 Y_{-1-p_o \over 2}(z)Y_{1-p_o \over 2}(z)  \\[3mm]
&\quad +(c_J c_Y^* + c_J^* c_Y)
\Bigl( J_{-1-p_o \over 2}(z) Y_{1-p_o \over 2}(z)% \\[3mm] 
%&\qquad \qquad \qquad \quad \quad
+ J_{1-p_o \over 2}(z) Y_{-1-p_o \over 2}(z) \Bigr)
 \biggr]^2 \ .  
\end{array}                                             \eqno(5.4)
$$

Now depending on the  range of $p_o$,   this quantum system is either  dominated by
 quantum fluctuations, or the quantum fluctuations are suppressed and it is represented
 by a classical geometry.  
To analyze this range, we first define,  
$$
\nu_1={1-p_o \over 2 } , \quad  \nu_2={-1-p_o \over 2} \ ; \quad  \nu_1 = \nu_2 + 1. 
                                                                 \eqno(5.5)
$$
Now we can perform a   similar analysis to the one done in the previous section. 
Thus, we can analyze various case for this system. 

Let us start by considering  $\nu_1 = 0$ or $\nu_2 = 0$. 
We observe that for  $p_o = -1$, we can write   
$$
\Delta \psi \cdot \Delta \pi \rightarrow \infty \quad  
( z \rightarrow 0 ).                                       \eqno(5.6)
$$
Now let us also consider the case $\nu_1 >0, $ and $\ \nu_2 >0$. 
For  $p_o < -1$, we obtain   
$$
\Delta \psi \cdot \Delta \pi \rightarrow \infty \quad  
( z \rightarrow 0 ) .                                       \eqno(5.7)
$$
For the $\nu_1 >0, $ and $ \ \nu_2 <0$, we observe that when  when  $-1 < p_o < 1$, 
$$
\Delta \psi \cdot \Delta \pi \rightarrow O(1) \quad  
( z \rightarrow 0 ) .                                       \eqno(5.8)
$$
Now for the case $\nu_1 <0 , \ \nu_2 <0$, when   $p_o > 1$, we obtain 
$$
\Delta \psi \cdot \Delta \pi \rightarrow \infty \quad  
( z \rightarrow 0 ).                                       \eqno(5.9)
$$ 

Summarizing above discussions ,  we obtain that 
$$
{\rm when} \ \  p_o > 1 \  {\rm or} \ p_o \leq -1 \ , \quad 
\Delta \psi \cdot \Delta \pi \rightarrow \infty \quad  ( z \rightarrow 0 ) ,                                                                 
                                                                                             \eqno(5.10)
$$
which means that when $p_o > 1$ or $p_o \leq -1$ the quantum fluctuations  dominate 
the universe at the early times.  
On the other hand we obtain that   
$$
{\rm when} \ -1 < p_o < 1 \ , \quad 
\Delta \psi \cdot \Delta \pi \rightarrow O(1) \quad  ( z \rightarrow 0 ) ,                                                                   
                                                                                             \eqno(5.11)
$$
which means that when $-1 < p_o < 1$ universe can become classical at the early times. 
Since we expect that the quantum fluctuations dominate the universe at the early times, 
$p_o > 1$ or $p_o \leq -1$ is desirable. 
Note that we have also assumed $p_o \neq 1$ , and above result is consistent with 
Ref.  \cite{OhkuwaFaizalEzawa1}.

Comparing this section and previous section, we find that in both models there exist the
 common ranges for  physically desirable $p_o$, that is from Eqs. (4.26) and (5.10) 
$$
{\rm when} \ \  p_o > 1 \  {\rm or} \ p_o \leq -5 \ , \quad 
\Delta \psi \cdot \Delta \pi \rightarrow \infty \quad  ( a \rightarrow 0 ) ,                                                                  
                                                                                            \eqno(5.12)
$$
which means that when $p_o > 1$ or $p_o \leq -5$ the quantum fluctuations  dominate 
the universe at the early times, $a \rightarrow 0$.  
Since these ranges of $p_o$ are very wide, we could expect that there might exist some
 model  independent  desirable operator ordering parameter  $p_o$ in the
 Wheeler-DeWitt equation.     
Note that, since our analysis is based on the assumption $p_o \neq 1$, there remains 
the possibility that $p_o = 1$  might be also the model  independent  desirable
 operator ordering parameter.

\section{Conclusion}\label{d}
In this paper, we have analyzed the creation of universe using third quantization. 
At the beginning of the  universe, the geometry of the universe  is dominated by quantum
 fluctuations. 
These fluctuations are suppressed as this universe evolves, resulting in  a classical
 geometry of our universe. 
We have used these two physical requirements to constraint the range of factor ordering
 for two different cosmological models. 
It was observed that both these cosmological models satisfy the desired evolution
 only for the common  ranges of $p_o$,  
$
{\rm when} \ \  p_o > 1 \  {\rm or} \ p_o \leq -5 \ , \quad 
\Delta \psi \cdot \Delta \pi \rightarrow \infty \quad  
( a \rightarrow 0 ).                                      
$
Thus, it seems that for the values    $p_o > 1$ or $p_o \leq -5$ the quantum fluctuations 
 dominate initial state of the universe $a \rightarrow 0$, and a classical geometry will
 form at later stages of the evolution of the universe. 
 It may be noted that as we have obtained the very wide common ranges for the desirable 
 operator ordering parameter    $p_o$  for two different cosmological models, it indicates
 that there is a possibility that there exists some desirable $p_o$ which is independent of the
 specifics details of a cosmological model. 
However, it would be important to analyze many other different cosmological 
 models to verify the model independence of this  value. 
Our analysis is based on the assumption $p_o \neq 1$, so it is possible that  $p_o = 1$
  might also be a  valid value for the operator ordering parameter. 
  
  It may be noted that the third quantization has been generalized to loop quantum
 gravity, and this has led to the development of group field theory
 \cite{gft12}-\cite{gft14}, and group field cosmology  \cite{gfc12}-\cite{gfc14}. It would be
 interesting to generalize the results of this paper to these third quantized models of loop
 quantum gravity.
 Furthermore, the third quantization of string theory has also been used to study the
 creation of a pair of universes from string vacuum state \cite{st}. 
It would be interesting to use the formalism developed in this paper to analyze the
 creation of universe using string theoretical solutions. 
We would also like to point out that the third quantization Horava-Lifshitz gravity  has
 also been discussed \cite{3}-\cite{4}.  
It would be interesting to analyze the  operator ordering ambiguity for such a cosmological 
 model. 
It may be noted as this is an non-trivial modification of gravity, if we obtain  similar ranges
for the  values of the operator ordering parameter, then this would be a strong
 indication of the  existence of the model independent operator ordering parameter.

 \end{document}